# Improved Efficiency of Photoconductive THz Emitters by Increasing the Effective Contact Length of Electrodes

Abhishek Singh[1], Harshad Surdi[1], V. V. Nikesh[1], S. S. Prabhu[1,a] and G. H. Döhler[2]

[1]Tata Institute of Fundamental Research, Homi Bhabha Road, Mumbai 400005, India

[2] Max-Planck Institute, University of Erlangen, Erlangen, Germany

**Abstract:** We study the effect of a surface modification at the interface between metallic electrodes and semiconducting substrate in Semi-Insulating GaAs (SI-GaAs) based photoconductive emitters (PCE) on the emission of Tera-Hertz (THz) radiation. We partially etch out 500 nm thick layer of SI-GaAs in grating like pattern with various periods before the contact deposition. By depositing the electrodes on the patterned surface, the electrodes follow the contour of the grating period. This increases the effective contact length of the electrodes per unit area of the active regions on the PCE. The maxima of the electric field amplitude of the THz pulses emitted from the patterned surface are enhanced by up to more than a factor 2 as compared to an un-patterned surface. We attribute this increase to the increase of the effective contact length of the electrode due to surface patterning.

Improving the emission efficiency of THz sources has been a major research goal for several years. There is considerable interest in creating high power and high amplitude electric field THz radiation due to its potential applications in spectroscopy and imaging[1]. However, the methods of enhancing

---

[a] Author to whom correspondence should be addressed. Electronic Mail: prabhu@tifr.res.in

THz radiation by non-linear optical and electronic techniques have several limitations. It is well established that for laser pulse energies in the nano-Joules range it is not advisable to use a non-linear optical method of THz generation[2,3], rather it is advisable to use photo-conductive antennas. Due to the recent advances in the field of ultra-short pulse lasers, compact high repetition rate laser sources have become easily available. However, as their pulse energies are still relatively modest, for compact photo-conductive THz sources, providing sufficiently large power, a high conversion efficiency, is highly desirable. Many designs of THz photo-conductive emitters (PCEs) have been tested to increase the overall THz power emission[4]. Several attempts have been made recently using plasmonic structures[5,6]. These include modifying the structure of the electrodes of the antenna[7], modification of the area between the two electrodes[5] or adding Silver nano-particles in the area between the two electrodes[8]. Since most of the THz emission from PCEs comes from the region very close to the anode[7,9], it is expected that the THz emission will increase, if we can enhance carrier generation and collection near the electrode. In this paper we have increased the fraction of photo-generated carriers near the electrode by increasing the effective length of the electrodes by patterning the substrate. In Figure 1(a) we show the schematic diagram of the PCE with two Au-Ge pads. The pads are separated by 25 microns. We etch several rectangular grating patterns on SI-GaAs substrate with different periods before contact deposition. As shown in Figure 1(c) we increased the total path covered by electrodes without increasing the apparent length of the electrodes, by etching the substrate surface periodically in rectangular grooves of 500 nm depth along the electrode length before Au-Ge electrode deposition. The depth of 500 nm has been kept the same for all etched patterns. The periods are 1 μm ($R_1$), 2μm ($R_2$) and 3μm ($R_3$). Un-patterned region is labeled as $R_0$. In the case of the electrodes on the flat SI-GaAs surface ($R_0$), the electric field lines between the two electrodes follow curved paths as we go deeper inside the crystal. As expected, the electric field strength decreases as we go deeper inside the crystal. Since the 800 nm laser pulses used for charge carrier generation penetrate only 1-2 μm into the SI-GaAs, the decrease in the electric field will not be very significant up to that depth but still it can't be ignored whereas, in the case of patterned regions, the un-etched stripes of SI-GaAs will exhibit relatively straight field lines and hence with increased magnitude.

To study the electric field distribution inside the crystal material between the two electrodes, we have done simulations using commercial software COMSOL. In Figure 1(c) we show the simulation results for the patterned and un-patterned SI-GaAs surface. Simulations were done for electrodes on a flat SI-GaAs surface ($R_0$), and on surfaces with depth modulation of 500 nm with periods of 1.0 µm ($R_1$), 2.0 µm ($R_2$) and 3.0 µm ($R_3$). The result shown in Figure 1(b) represents the electric field strength distribution 200 nm below the surface of the un-etched crystal. Since in the patterned region, 500 nm deep alternate slabs of SI-GaAs have been etched out, those regions are left out in the shown result of electric field distribution. In the case of 25 µm electrode gap and 25 V applied bias, the surface modulation has increased the electric field near the electrodes to values considerably exceeding the average of 10 kV/cm. This is expected because these stripes are surrounded by electrodes along three faces. This increase in electric field is expected to be higher for closer electrodes. While there will be slight enhancement for the electric field in the un-etched slabs region, the field in the regions below etched out slabs will be similar to the regions below the usual un-patterned area. Over all emission of THz is expected to increase not only because of the slight increase in electric field but also from an increase in effective contact length of electrodes in contact with SI-GaAs, which will eventually increase the number of electrons generated in close vicinity of the anode. Since, most of the THz emission comes from carriers in the close vicinity of the anode[7,9], the increase of their number will result in an increase the overall THz emission.

To avoid the sample to sample fabrication variations, all three regions $R_1$, $R_2$ and $R_3$ were arranged side by side on a single SI-GaAs substrate using standard Electron Beam Lithography and Reactive Ion Etching (RIE). To study the surface morphology of the etched regions Atomic Force Microscopy (AFM) was used. In Figure 2 we show the AFM images of 0.5 µm ($R_1$) and 1.0 µm ($R_2$) wide grating lines. The images show that the depth of the etched region of SI-GaAs surface corresponds rather well to the desired value of 500 nm. In the fabrication process, the surface was etched first and then two 3mm long, 50 µm wide and 200 nm thick AuGe electrodes, separated by 25µm gap-distance, were deposited on the SI-GaAs surface in such a way that it was covering all three patterned regions $R_1$, $R_2$, $R_3$ and some un-patterned region $R_0$. For all patterns the grating lines are perpendicular to the electrode direction. Standard lift-off photo-lithography technique was used to deposit the electrode contacts.

Four regions of the PCE source, $R_0$, $R_1$, $R_2$ and $R_3$, were gated one by one with 800nm, 10fs, 76MHz laser pulses and the incident optical beam was focused using an off-axis parabolic mirror of 2.0 inch focal length giving a focused beam spot of ~50 µm. The emitted THz radiation was detected using a standard THz-time domain spectroscopy (THz-TDS) set up with ZnTe <110> crystal for electro-optic detection. The source was electronically chopped using a square-wave voltage pulse waveform generator, operated at a frequency of 27 KHz. We recorded the time domain THz signal emitted from the PCEs. They were analyzed in terms of maximum amplitude of the emitted THz electric field and, by Fourier transformation, regarding the frequency distribution of the emitted THz radiation. Source was tested without any THz collimating silicon lens on the other side of source to avoid the THz signal variations coming from its misalignment.

The THz emission of the different regions $R_0$, $R_1$, $R_2$ and $R_3$ of the antenna was studied as a function of applied bias voltage at two fixed excitation powers of 50 mW and 100 mW. In Figure 3 we depict for both excitation powers the observed THz field maxima as a function of applied voltage for all the four regions $R_0$ – $R_3$. The measured THz peak scales (a.u.) shown on left axis are different for 50 mW and 100 mW case. In the patterned regions a rather strong increase of the THz field maxima, exceeding a factor of two for the region with the shortest period, is observed. This increase can be understood by the following (simplifying) consideration. The absorption of the laser pulse yields an electron-hole plasma in the illuminated area not shadowed by the electrodes. The density of this plasma decreases exponentially with a characteristic length corresponding to the inverse of the absorption constant $\alpha$ ($\alpha^{-1} \approx 1000$ nm for $\lambda = 800$ nm in GaAs). On the scale of our modulation depth of 500 nm the density of this plasma can be considered as nearly uniform. We will see below that only carriers within a small distance $s$ from the contact are contributing to the THz pulse. If this distance $s$ is even small compared to the characteristic lengths $a$, $b$ and $c$ of the surface patterning (see Fig. 1(c)) this implies that the photo-generated carriers within a distance $s$ from the (vertical) sidewalls of length $c$ are contributing to the THz signal by drifting towards the neighboring sidewall while those within a distance $s$ from the horizontal contact regions of length $a$ at the top of the stripes, or of length $b$ at the bottom of the grooves are drifting towards these contact regions. Therefore, the THz current and the THz field are scaling with the "effective contact length" per period, $l^{eff} = a + b + 2c$, instead of the period $l = a + b$. Compared with the un-patterned region $R_0$ this yields an enhancement factor $\eta = l^{eff} / l = (a + b + 2c) / (a + b)$. For the patterns $R_1$, $R_2$ and $R_3$ in Fig. 1(c) we obtain $\eta_1 = 2$, $\eta_2 = 1.5$ and $\eta_3 = 1.33$, respectively.

An upper limit for the value of $s$ can be obtained from an estimate of the distance that photo-generated carriers can drift during a THz pulse under the influence of the DC electric field in the sample. The value of $s$ is much larger for the (light) electrons compared with the holes. The transport can be considered as quasi-ballistic as the fields are in the range $> 5$ kV/cm and the pulse duration will be in the order of 1 ps or less. For this scenario Monte Carlo simulations of ballistic electron transport for GaAs yield distances $s < 200$ nm and pulse duration $< 1$ ps (see, for instance, Figure 7 in Ref.11). Thus, $s << (a, b, c)$ represents, at least, a reasonable approximation for our system. Interestingly, the results for the enhancement of the THz fields indicate that the enhancements observed even exceed values expected due to the increased effective contact lengths. This may be due to stronger field enhancement near the contacts in the patterned structures. Also, the enhancement factors exhibit a significant dependence on the laser power and on the applied voltage, which needs further studies.

Finally, the FFT spectrum of the time-domain THz signal is shown in Figure 4. The shape of the Fourier spectrum doesn't show any difference between the signals from un-patterned ($R_0$) and patterned regions ($R_1$, $R_2$ and $R_3$). This result is consistent with the expectation that patterning only increases the number, but does not affect the kinetics of the carriers collected at the contacts, provided that $s \ll a,b$ and $c$.

In summary, we have performed a comparative study of photoconductive emitters (PCEs) with regular electrodes on a flat semiconductor surface (region $R_0$) and electrodes on periodically modulated surfaces (patterned regions $R_1$, $R_2$ and $R_3$). The ratio of peak amplitudes of emitted THz pulses from patterned to un-patterned region of antenna shows maximum scales roughly with the "effective contact length", which includes the vertical contact areas. An enhancement even exceeding the expected factor of 2 has been observed in patterned samples with effective contact length twice the horizontal contact length.

We thank A.V. Gopal for his critical comments on the MS.

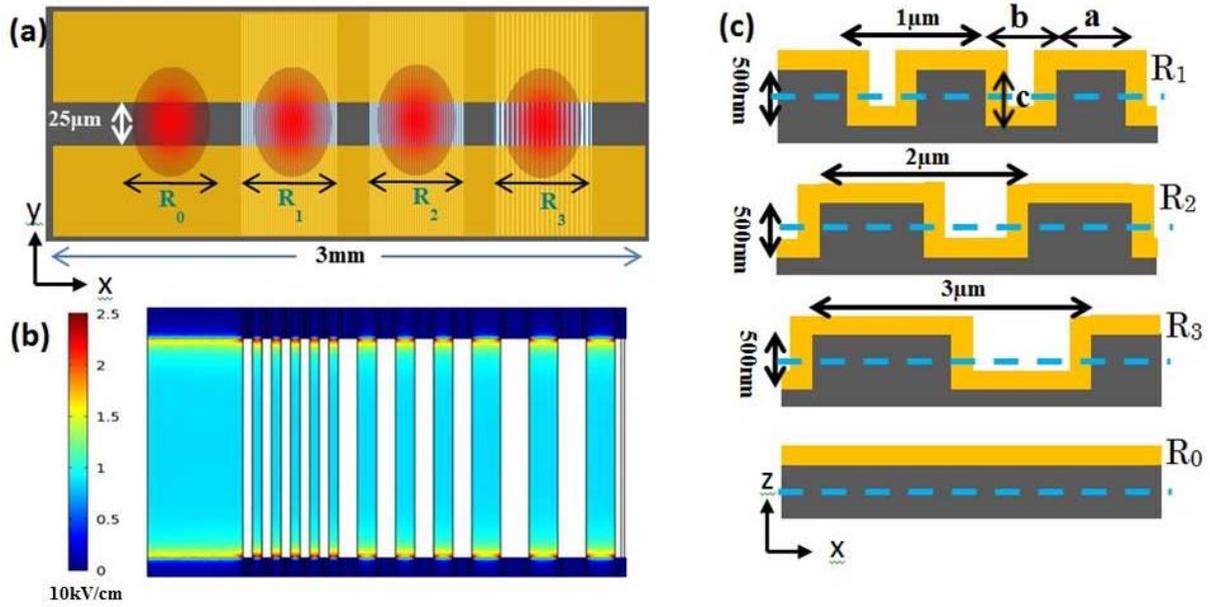

**FIG. 1:** (a) The schematic diagram of PCE THz source; red color are showing the laser excitation spot at regions with different surface morphologies. (b) Simulated Electric field in a plane 200 nm deep inside the SI-GaAs at different regions $R_0$, $R_1$, $R_2$ and $R_3$ for applied bias of 25 V across 25 μm gap. (c) Cross sections of electrodes over the SI-GaAs slabs at different regions $R_1$, $R_2$ $R_3$ and $R_0$ from top to bottom. Dotted lines are showing the plane of simulated electric field.

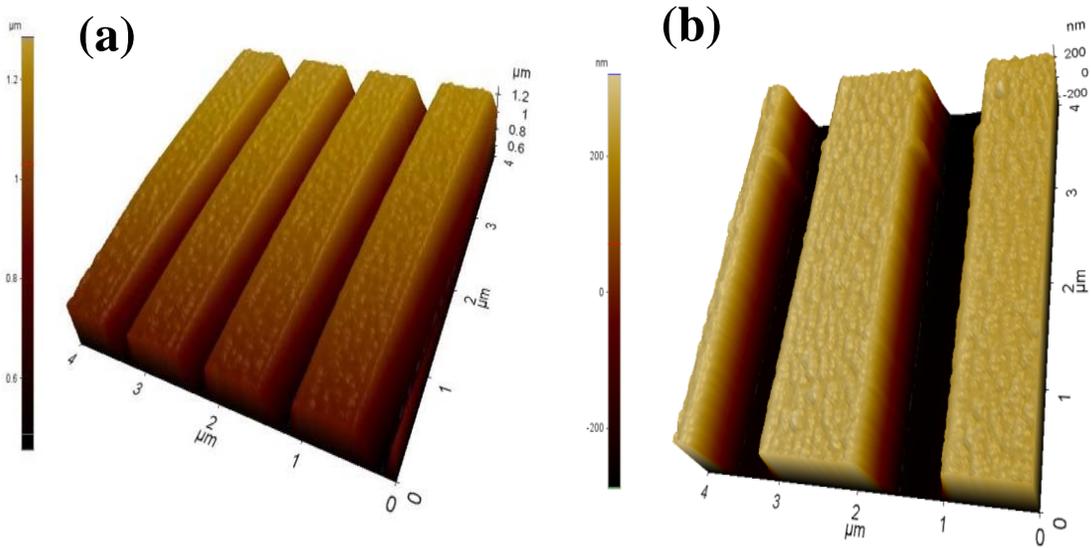

**FIG. 2:** AFM Images of region $R_1$(a) and $R_2$(b) having grating periods of 1μm and 2μm respectively.

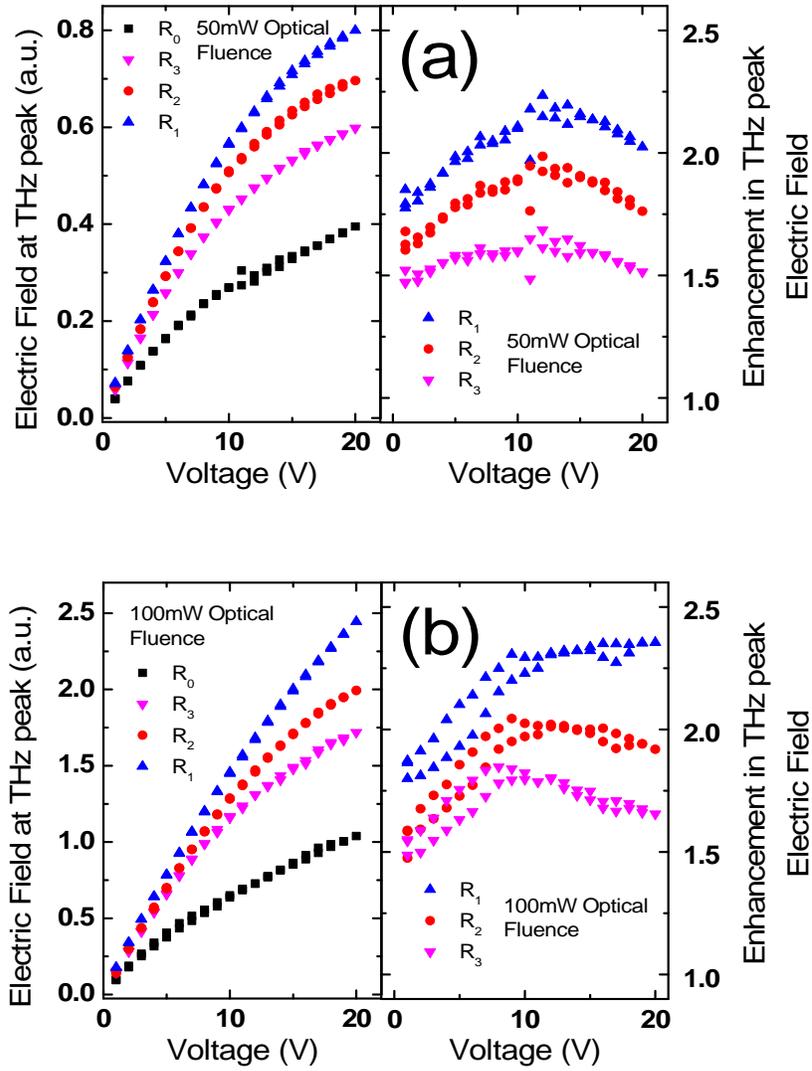

**FIG. 3:** Variation of THz peak electric field with applied bias (left) and relative enhancement in THz peak electric field (right) at different patterned regions for the optical fluence of **(a)**50 mW and **(b)**100 mW. Voltage was varied from zero to 20 V and then back to zero from 20 V. The measured THz peak scales (a.u.) shown on left axis are different in (a) and (b) cases.

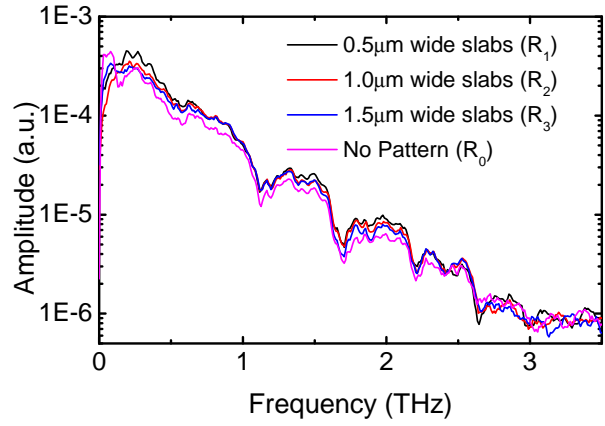

**FIG. 4:** FFT Spectrum of THz pulses emitted from different regions of PCE.